# Green's Function Decomposition for a New Rewriting of Inverse Scattering Equations

Martina T. Bevacqua and Tommaso Isernia

*Abstract*—In this paper, a new inversion model for two-dimensional microwave imaging is introduced by means of a convenient rewriting of the usual Lippmann–Schwinger integral scattering equation. Such model is derived by decomposing the Green's function and the corresponding internal radiation operator in two different contributions. In fact, one of them can be easily computed from the collected scattered data. In case of lossless backgrounds, the resulting model turns out to be more convenient than the traditional one, as it exhibits a lower degree of nonlinearity with respect to parameters embedding the unknown dielectric characteristics. This interesting property suggests its exploitation in the solution of the inverse scattering problem. The achievable performances are tested by comparing the proposed model with the usual one based on the Lippman-Schwinger equation in both cases of linearly approximated and full non-linear frameworks. Both numerical and experimental data are considered.

*Index Terms*—Born approximation method; Contrast Source Inversion; Green's function, inverse scattering; microwave imaging; non-linearity; radiating currents.

## I. INTRODUCTION

THE development of effective and accurate techniques for the solution of inverse scattering problems [1]-[2] plays a pivotal role in very many as well as relevant microwave applications, such as biomedical imaging, subsurface prospecting and non-destructive testing [3]-[6]. The adopted mathematical models for describing the electromagnetic scattering phenomena significantly affect the reliability and the achievable performance [7]-[9]. As a consequence, a large attention has been paid on the analysis and formulation of new scattering models, able to enlarge applicability and accuracy of the adopted inversion techniques. In particular, scattering models are of interest such as reduce as much as possible the non-linearity of inverse scattering problems, and hence counteract the false solutions problem [10].

With respect to the two-dimensional geometry and scalar field case, some analyses have been performed to quantify the Degree of Non-Linearity (DNL) of the relationship between the unknown permittivity profile and the scattered field as a measure of the difficulty of inverse scattering problems. In particular, the DNL has been evaluated depending on the scatterer size as well as the maximum magnitude of the unknown permittivity profile [7]. The larger the permittivity and/or the target dimensions, the higher the nonlinearity of inverse scattering problems. The non-linearity of inverse scattering problems can be mathematically explored through the 'state' equation, that is the Lippmann–Schwinger integral equation, which relates the induced current/total field inside the investigation domain to the unknow target properties. As such, the DNL has been proved to be strictly connected to the norm of the radiation operator adopted in the state equation, which takes into account multiple scattering effects [7].

To reduce the nonlinearity, in [8] a convenient rewriting of the scattering equations has been derived from the traditional ones by taking advantage from the peaked behavior of the Green's function in lossy media and without adopting any approximation. Such rewriting of the equation modelling the scattering phenomena has allowed a reduction of the DNL of the problem. The approach has been also proved to be useful in case of lossless backgrounds [9].

In the same spirit, in [11] a family of new integral equations, which are transformed from the original Lippmann–Schwinger integral equation and wherein the model [8],[9] can be seen as a specific case, has been introduced. In such new models, issues arising from non-linearity are effectively alleviated, and, again, no approximation is involved.

Encouraged from the results in [8],[9],[11], in this paper a new mathematical model is introduced to solve inverse scattering problems in lossless and homogenous background medium which is based on a convenient decomposition of the Green's function and the corresponding internal radiation operator. In fact, one of the two resulting integrals turns out to be easily computable from the scattered field, as preliminary discussed in [12]. Notably, such a contribution, by virtue of the results in [13], is indeed related to (a part of) the radiating currents [14],[15] induced inside the unknown target. Such a circumstance suggests some connections with the Subspace Optimization Method (SOM) [16] and a non-iterative method introduced by X. Chen et al. in [17], but a deepening of these connections is outside the scope of the present paper. A Green's function decomposition is also used in [18], wherein a data-driven linearized approach is derived by assuming as an

Martina T. Bevacqua and Tommaso Isernia are with the Dipartimento di Ingegneria dell'Informazione, delle Infrastrutture e dell'Energia Sostenibile (DIIES), Università Mediterranea di Reggio Calabria, 89124 Reggio Calabria, Italy, also with the Institute for Electromagnetic Sensing of the Environment, National Research Council of Italy (CNR-IREA), 80124 Naples, Italy (e-mail: martina.bevacqua@unirc.it; tommaso.isernia@unirc.it) and also with Consorzio Nazionale Interuniversitario per le Telecomunicazioni (CNIT) Viale G.P. Usberti, 181/A Pal.3 - 43124 Parma, Italy.

auxiliary unknown the field which would be scattered from the target under the Born approximation (for the internal field), which is also very different from what follows.

As discussed in the following, the capability to evaluate one of the two integrals resulting from the Green's function decomposition allows to get an equation which has exactly the same structure as the original Lippman Schwinger equations, but with a different and somehow more convenient integral operator. In fact, the introduced model can be proved to exhibit a lower DNL with respect to parameters embedding dielectric characteristics as compared to the traditional scattering equations. As a consequence, the use of the proposed model implies interesting advantages in term of convergence and accuracy of the corresponding inversion procedures. Such benefits are proved in the following both in linear and non-linear frameworks. In particular, we introduce a new linear inversion method, able to enlarge the range of validity of the classical BA [19], and a modified version of the well-known contrast source inversion (CSI) method [20].

The paper is organized as it follows. In Section II, the traditional equations modelling inverse scattering problems are reported and the concept of degree of non-linearity and its relevance are recalled. Section III introduces the new scattering model herein proposed. Finally, in Sections IV and V some numerical analyses are performed to test the proposed model within linear and non-linear frameworks, respectively. In particular, both simulated and experimental single frequency data are processed. Conclusions follow. Throughout the paper, the case of scalar fields and two-dimensional geometry is considered and a time harmonic factor $exp\{j\omega t\}$ is assumed and dropped.

## II.  MATH OF THE PROBLEM

### A.  The classical scattering model

Let us consider one or more unknown dielectric scatterers in the investigated domain D. Let $\Sigma$ denote their support and $\varepsilon_b(\underline{r})$ and $\varepsilon_x(\underline{r})$ the complex permittivity of the background medium and the unknown targets, respectively, with $\underline{r} = (x, y)$. Let us probe D with some transmitting antennas located in $\underline{r}_t \in \Gamma$ outside D. The classical model describing the scattering phenomenon for the generic incident direction $v$ corresponding to each $\underline{r}_t$ positions, is composed by two integral equations, the data and state equations, that are respectively: [1]:

$$E_s^v(\underline{r}_m) = \int_D G_b(\underline{r}_m, \underline{r}') \chi(\underline{r}) E_t^v(\underline{r}') dr' = A_e(W^v)$$
(1)

$$W^v(\underline{r}) = \chi(\underline{r}) E_i^v(\underline{r})$$
$$+ \chi(\underline{r}) \int_D G_b(\underline{r}, \underline{r}') W^v(\underline{r}') dr'$$
$$= \chi(\underline{r}) E_i^v(\underline{r}) + \chi(\underline{r}) A_i[W^v(\underline{r})]$$
(2)

---

[1] The background medium is assumed to be homogeneous and lossless.
[2] Note that $\|A_i X\| < 1$ is a sufficient condition for writing the series (3).

where $\chi(\underline{r}) = \frac{\varepsilon_x(\underline{r})}{\varepsilon_b(\underline{r})} - 1$ is the contrast function which encodes the electromagnetic properties of the unknown objects, $E_s^v(\underline{r}_m)$ is the scattered field measured by different receivers located in $\underline{r}_m \in \Gamma$ outside D. $W^v = \chi E_t^v$, $E_t^v$ and $E_i^v$ are, respectively, the induced currents, the total field and incident electric field in D. $A_e$ and $A_i$ are short notations of the external and internal radiation operators, respectively. Finally, $G_b(\underline{r}, \underline{r}') = -\frac{j}{4} k_b^2 H_0^2(k_b|\underline{r} - \underline{r}'|)$ is the Green's function pertaining to the background medium[1], being $H_0^2$ the zero order and second kind Hankel function and $k_b = \omega\sqrt{\mu_b \varepsilon_b}$ the wavenumber in the host medium.

The equation (2) is the Lippman Schwinger equation and relates the induced currents/total field inside D to the contrast function $\chi$ [1]. For the sake of brevity as well as for emphasizing the differences with the one proposed in the following, let us identify the model (1)-(2) as the H02 model.

### B.  A measure of the 'degree of non-linearity'

In order to establish the complexity and difficulty of the inverse scattering problem at hand, a key role is played by the norm of the operator $A_i X$ involved in the state equation (2), (wherein $X(\cdot)$ is the diagonal operator that gives the product $\chi$ times $(\cdot)$). If the $l_2$-norm $\|A_i X\|$ is lower than 1, the inverse operator $(I - A_i X)^{-1}$, which relates the total field to the incident one, can be expanded in a Neumann series[2] as follows [7]:

$$(I - A_i X)^{-1} = I + A_i X + (A_i X)^2 + \cdots + (A_i X)^n + \cdots$$
(3)

wherein $I$ is the identity operator. From the above series, one can infer that the overall DNL of a given scattering problem increases with the norm of the operator $A_i X$. In fact, one can foresee what is the minimum number of terms required to achieve a given approximation accuracy [7]-[9]. For example, if $\|A_i X\| \ll 1$, one can just consider the first term and a linear relationship holds true between data and unknowns [19]. On the other hand, by truncating the series at the second term, the scattered field can be expressed as a quadratic function of the contrast [21]. If $\|A_i X\| < 1$, for every given accuracy a polynomial relationship holds true between data and unknowns of the inverse problem, and the order of the polynomial depends on how large is $\|A_i X\|$. If $\|A_i X\| > 1$ a non-polynomial relationship instead holds true between data and unknowns. As a consequence of the above, the larger $\|A_i X\|$ is, the larger are the DNL and the overall difficulty of the problem [7]-[9]. In fact, the cost functional whose global minimum defines the solution of the inverse problem is a polynomial with double order with respect to the one defined from the series in (3). Hence, the value of $\|A_i X\|$ also gives a quantitative information on the possible number of local minima (corresponding to false solutions) of the cost functional at hand.

Then, understanding the factors affecting $\|A_i X\|$ is fundamental in order to keep under control the occurrence of false solutions. In this respect, note that by applying the Schwarz's inequality, an upper bound to $\|A_i X\|$ can be obtained as:

$$\|A_i X\| < \|X\| \|A_i\| \tag{4}$$

In such a way, the role played by the integral operator $A_i$, which only depends on the kernel and the domain of the integral operator, is separated by the one played by contrast function $\chi$, which accounts for the properties of the unknown targets. Let us stress that for a fixed contrast function, the non-linearity of the problem depends indeed on the properties of the integral operator adopted in the state equation.

Then, by using (4), a (sufficient) condition for the applicability of the series (3), as well as additional information about its DNL, can be gained by separately investigating $\|X\|$ and $\|A_i\|$.

## III. A NEW SCATTERING MODEL FROM GREEN'S FUNCTION DECOMPOSITION

The Green function is the solution of the wave equation for a point source. It is the impulsive response of the system. Indeed, if the Green function is known, the solution of the wave equation due to a general source can be deduced thanks to the linear superposition [22].

According to [23], the Green's function is the superposition of a homogenous and inhomogeneous components. Both of them include propagating waves, while the homogeneous parts contain only propagating ones. Moreover, the singularity of the Green function is contained completely in the inhomogeneous part.

According to the properties of the Hankel function [24], the Green's function in (1)-(2) can be decomposed in two terms, containing respectively the zero order Bessel functions of the first kind $J_0$ and second kind $Y_0$, i.e.:

$$G_b(\underline{r}, \underline{r}') = -\frac{jk_b^2}{4} J_0(k_b|\underline{r} - \underline{r}'|) - \frac{k_b^2}{4} Y_0(k_b|\underline{r} - \underline{r}'|) \tag{5}$$

The first and second terms represent, respectively, the homogenous and inhomogeneous components of the Green's function, as in [23]. The Bessel function $J_0$ and $Y_0$ in the above equation exhibit different properties both in the spatial and spectral domains [24]. For instance, unlike $J_0$ which is a continuous function of $\underline{r} - \underline{r}'$, $Y_0$ exhibits a singularity in $\underline{r} = \underline{r}'$. Moreover, $J_0$ has a spectral content only concentrated in the circle of radius $k_b$. On the other side, the inhomogeneous part of the Green function, that is $Y_0$, has positive spectral components outside the circle, zero on it and negative inside it [23].

Then, by exploiting the decomposition of the Green's function in homogeneous and inhomogeneous parts, the internal radiation operator $A_i$ can be split into the sum of two new integral internal operators $A_i^{J_0}$ and $A_i^{Y_0}$, i.e.:

$$\begin{aligned} A_i(W^v) &= -j\frac{k_b^2}{4} \int_D J_0(k_b|\underline{r} - \underline{r}'|) W^v(\underline{r}') d\underline{r}' \\ &\quad - \frac{k_b^2}{4} \int_D Y_0(k_b|\underline{r} - \underline{r}'|) W^v(\underline{r}') d\underline{r}' \\ &= A_i^{J_0}(W^v) + A_i^{Y_0}(W^v) \end{aligned} \tag{6}$$

As discussed in the following subsection, the first contribution at the left hand of the equality (6), which in the following is referred as $E_{J_0}{}^v$, can be easily computed from the collected scattered data.

### A. On the meaning and computation of $E_{J_0}{}^v$

The first integral in (6) can be seen as the convolution product between the induced contrast sources and the relevant zero order Bessel function $J_0$. Hence, due to the spectral properties of the Bessel function $J_0$, the first convolution product in (6) extracts the spectral component of the currents located on the circle of radius $k_b$. This is a very interesting circumstance as, by virtue of results in [14],[15],[25],[26], radiating sources oscillate indeed at those frequencies (but for subtleties related to the finiteness of sources). As a consequence, the first part of the decomposition in (6) can be related to the radiating currents [13]. In particular, the first integral can be interpreted as the main contribution of the radiating currents, i.e. the one lying on the circle of radius $k_b$ in the spectral domain, while the remaining part is still present in the second term $A_i^{Y_0}$ [23],[25] (see [13] for more details). As a consequence, the first integral in (6) can be understood as the contribution to the total field inside D by the main spectral component of the radiating currents, which are indeed peaked along the circle of radius $k_b$ in the spectral domain [23],[25].

The natural question then arises on how one can compute $E_{J_0}{}^v$. Notably, the radiating currents $W_{rad}^v$, that are the part of the currents responsible of scattering phenomenon, can be computed from the data by solving the inverse source problem described by the data equation (1). To this end, due to the severe ill-posedness, a regularization technique has to be adopted, such as the Truncated Singular Value Decomposition (TSVD) or the Tikhonov regularization [2]. In applying the desired regularization technique, one is also retrieving the spectral components of the radiating currents not lying on the circle $k_b$. Then, in order to extract just the spectral component located on the circle, which are of interest in the identity (6), one can use the $J_0$-filter, that is one can evaluate the term $A_i^{J_0}(W_{rad}^v)$. In fact, as discussed above, the spectral properties of $J_0$ are such to filter out the contributions located outside $k = k_b$.

A second and even more interesting possibility to compute $E_{J_0}{}^v$, which is adopted in the following numerical tests, takes advantage from the results in [13],[27]. In particular, according to these latter, the first integral in equation (6) can be easily computed from the data as follows:

$$\int_D J_0(k_b|\underline{r} - \underline{r}'|) W(\underline{r}') d\underline{r}' = \int_\Gamma E_s^v(\underline{r}_m) K^{TM}(\underline{r}_m, \underline{r}) d\underline{r}_m$$

wherein:

$$(7)$$

$$K^{TM}(\underline{r}_m, \underline{r}) = \frac{1}{2\pi|\underline{r}_m|} \sum_{n=-\infty}^{+\infty} \frac{J_n(k_b|\underline{r}|)}{H_n^2(k_b|\underline{r}_m|)} e^{jn(\hat{r} - \hat{r}_m)}$$

$$(8)$$

$n$ is the order of the Bessel and Hankel functions, and $\Gamma$ is a closed curve of radius $|\underline{r}_m|$, wherein the measurements antennas are supposed to be located. The identity (7) represents the scalar product over the measurement domain $\Gamma$ of the measured data with the conjugate of $K^{TM}$. Note that, in evaluating this latter by Eq. (8), only the order $n$ belonging to the interval [$-k_b a$, $+k_b a$] can be considered, wherein $a$ denotes the radius of a ball that encloses D [28],[29].

Notably, in case of far field measurement configuration, the identity (7) reduces to:

$$\int_D J_0(k_b|\underline{r} - \underline{r}'|)\, W^v(\underline{r}')\, d\underline{r}' = \int_\Gamma E_s^\infty(\hat{r}_m, v)\, e^{jk_b \underline{r} \cdot \hat{r}_m}\, d\hat{r}_m$$

$$(9)$$

wherein $E_s^\infty$ is the far-field pattern and $\hat{r}_m$ identifies the direction of $\underline{r}_m$. The identity (9) simply represents the scalar product over the measurement domain $\Gamma$ of the measured far-field pattern with the test function $e^{-jk_r \cdot \hat{r}_m}$, that is nothing that the Green's function in far-field zone.

Hence, as long as the measurement probes surround the region under test and are located in a closed curve, the first contribution at the left hand of the equality (6), can be easily computed from the collected scattered data according to (7) or (9).

### B. The Y0 model

By taking into account the above circumstances, the Lippman-Schwinger scattering equation in (2) can be transformed into:

$$W^v(\underline{r}) = \chi(\underline{r})\hat{E}_i^v(\underline{r}) + \chi(\underline{r})A_i^{Y_0}(W^v)$$

$$(10)$$

wherein $\hat{E}_i^v(\underline{r}) = E_i^v(\underline{r}) - j\frac{k_b^2}{4}E_{J_0}{}^v(\underline{r})$. Together with (1), the state equation (10) identifies the new model, referred in the remainder of the paper as the Y0 model, wherein the fundamental quantities are again the induced currents $W$ and the contrast function $\chi$, and where the structure of the equation concerning the internal fields is identical to the one in H02 model, but where the integral internal operator $A_i$ has now been replaced by $A_i^{Y_0}$ and the incident field $E_i^v$ by a new known field $\hat{E}_i^v$. Note that, differently from the CS-EB model, introduced and discussed in [8],[9], the geometrical and electromagnetic properties of the targets are here still encoded in the contrast function $\chi(\underline{r})$. As a consequence, there is no need to adopt further procedure to extract the target features.

As equation (10) has exactly the same structure as the traditional 2D scalar integral equation (2), one can use the same solution strategies usually adopted to solve inverse scattering problems, such as for instance the BA or the CSI methods.

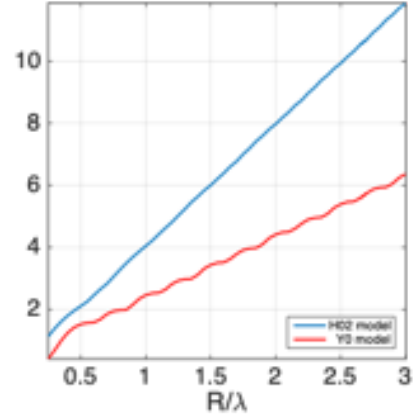

Figure 1. Universal plot as a function of the electrical size of the region under test: $\|A_i\|$ (blue line) and $\|A_i^{Y_0}\|$ (red line).

### C. Comparing Y0 and H02 models

The decomposition of the Green's function has involved the definition of a new integral operator $A_i^{Y_0}$. A comparison between the H02 model (1)-(2) and the herein proposed Y0 model (1)-(10) can be performed in term of DNL.

By virtue of the inequality (4), one can separately analyze the roles of the contrast profile and the relevant integral operator. In particular, one can compare the quantities at the right-hand side of (4) and of the corresponding inequality which holds true for Y0 model, i.e.,

$$\|A_i^{Y_0}X\| < \|X\|\|A_i^{Y_0}\|$$

$$(11)$$

As the factor $\|X\|$ is the same in the two inequalities (4) and (11), the DNL of the two formulations can be simply compared by only considering $\|A_i^{Y_0}\|$ and $\|A_i\|$.

To this end, a numerical analysis has been performed by considering that both operators depend only on the size (and shape) of D, whereas the electromagnetic characteristics of the (lossless) background medium can be taken into account by properly scaling the wavelength. Therefore, by following the same reasoning in [9] and by considering a circular domain D of radius R, it is possible to build up a plot of the norm of the two operators as a function of R/$\lambda_b$, where $\lambda_b$ is the wavelength in the background medium.

Figure 1 shows the universal plot of $\|A_i\|$ and $\|A_i^{Y_0}\|$, respectively. As can be seen, both norms are of course monotonically increasing functions of R/$\lambda_b$. Interestingly, one can notice that $\|A_i^{Y_0}\|$ is always lower than $\|A_i\|$. As such, for a fixed scattering problem, the new proposed model exhibits a lower DNL and, consequently, a lower number of false solutions.

### IV. NUMERICAL ASSESSMENT: LINEAR INVERSION

In order to test the new proposed model, a controlled assessment with both simulated and experimental data has been carried out by performing inversions within both a linear and (full) non-linear frameworks. In this section, a comparison between the results obtained by means of the classical BA within the H02 model and a new linear approximation derived from Y0 model has been performed.

In the standard BA, the inverse scattering problem is linearized by assuming the unknown total field D equal to the incident field. This hypothesis is fully satisfied only in case of weak scatterers, when $\|\chi A_i\| \ll 1$.

In the Y0 model, the incident field has been replaced with an equivalent one, which includes the contribution of a part of the radiating currents $E_{J_0}{}^v$. As such, a new linear approximation for the total field $E_t^v$ can be introduced as follow:

$$E_t^v(\underline{r}) \cong E_i^v(\underline{r}) - j\frac{k_b{}^2}{4}E_{J_0}{}^v(\underline{r})$$

(12)

The above approximation is valid when $\|\chi A_i^{Y_0}\| \ll 1$. As shown in figure 1, $\|A_i^{Y_0}\|$ is always lower than $\|A_i\|$, so that a wider range of applicability of approximation (12) is expected. By substituting this latter in the scattering model, the problem becomes linear, but it is still ill-posed, so that it has to be solved in a regularized sense. In the following examples, the TSVD regularization has been adopted [2]. In summary, the proposed linear inversion method derived from Y0 model, in the following referred as Y0-BA, is composed of the following steps:

- evaluation of $E_{J_0}{}^v$ from the data $E_s^v(\underline{r}_m)$ via the identity (7);
- evaluation of the approximated total field $E_t^v(\underline{r})$ by eq. (12);
- linearization of the problem in (1) and solution via TSVD regularization.

### A. Simulated data

In the following numerical analysis two different targets, embedded in free space, have been considered. Firstly, a homogeneous and lossless kite target has been positioned inside a square domain of side L = $\lambda_b$ and, following [29], 12 receivers and transmitters, modelled as line sources located on a circumference $\Gamma$ of radius R = 10 $\lambda_b$, have been considered. The scattered field data, simulated by means of a full wave forward solver based on the method of moments at the frequency of 300 MHz, have been corrupted with a random Gaussian noise with a given SNR = 30dB. Following [30], a number of cells $N_c$ equal to 60 × 60 has been used.

In order to compare the two models and prove the larger range of applicability of Y0-BA, different value of the contrast function have been considered, in particular, $\chi$=0.3, $\chi$=0.5, $\chi$=0.7 and $\chi$=1. The contrast profiles reconstructed by means of BA and Y0-BA are reported in Figures 2, 3 and 4. In addition, the figures also depict the profiles corresponding to the "ideal" case of known total field, that is, the contrast profile estimated when (1) is inverted by considering the exact total field, rather than the approximated one. While such a processing is obviously impossible in practice, it provides a benchmark, being the best possible result that can be achieved from the inversion of the data equation.

In Table I, the normalized mean square errors between the retrieved contrast function $\tilde{\chi}$ and the actual one $\chi$, defined as:

$$NMSE = \frac{\|\chi - \tilde{\chi}\|^2}{\|\chi\|^2}$$

(13)

are reported in order to quantitatively evaluate the obtained different performance.

Both Figures 2, 3 and 4 and Table I prove that the new proposed model is more convenient, as the corresponding linear approximation has a wider range of validity. Indeed, the Y0-BA allows to reach a lower NMSE than BA, and provided $\chi$ is not too large, is also able to exhibit performance similar to the ones obtained in the ideal case.

Then, as second numerical example, the well-known Austria profile with permittivity equal to 1.4 has been used as the ground-truth profile. The investigation domain of side L = 2$\lambda_b$ and discretized in $N_c$ = 64 × 64 cells has been investigated by means 26 receivers and transmitter located on $\Gamma$ with radius R = 13.3 $\lambda_b$. The data have been corrupted with a SNR=30 dB at the working frequency of 400 MHz. The results are reported in Figure 5. The means square errors are, respectively, 0.13 when the total field has been assumed exactly known, 0.40 when the new linear approximation derived from Y0 model has been adopted, and 0.89 when BA has been considered. As can be seen, despite the complexity of the Austria target, Y0-BA retrieves both the electromagnetic properties and shape of the target.

### B. Experimental data

In this subsection, the Fresnel targets, typically adopted to benchmark inverse scattering procedures, have considered, in particular:

- the *TwinDielTM* target [31], consisting of two circular dielectric cylinders with radius 1.5 cm and relative permittivity 3 ± 0.3;
- the *FoamDielIntTM* target, which is an inhomogeneous object, constituted by two nested circular cylinders an outer one made of foam (radius 40 mm, relative permittivity 1.45) that hosts another circular cylinder made of berylon (radius 15 mm and permittivity 3) [32].

The Fresnel data are collected in a partially limited aspect measurements configuration. More details about the targets and the measurement set-up can be found in [31],[32]. For the first target, the investigated area of 0.15 × 0.15 $m^2$ has been discretized in 64 × 64 cells, the working frequency has been selected equal to 4GHz, and a 72 × 36 multiview-multistatic data matrix has been processed. On the other hand, for the *FoamDielIntTM* target, the investigated area of 0.2 × 0.2 $m^2$ has been discretized in 78 × 78 cells and a 45 × 36 multiview-multistatic data matrix has been processed at the working frequency of 3GHz.

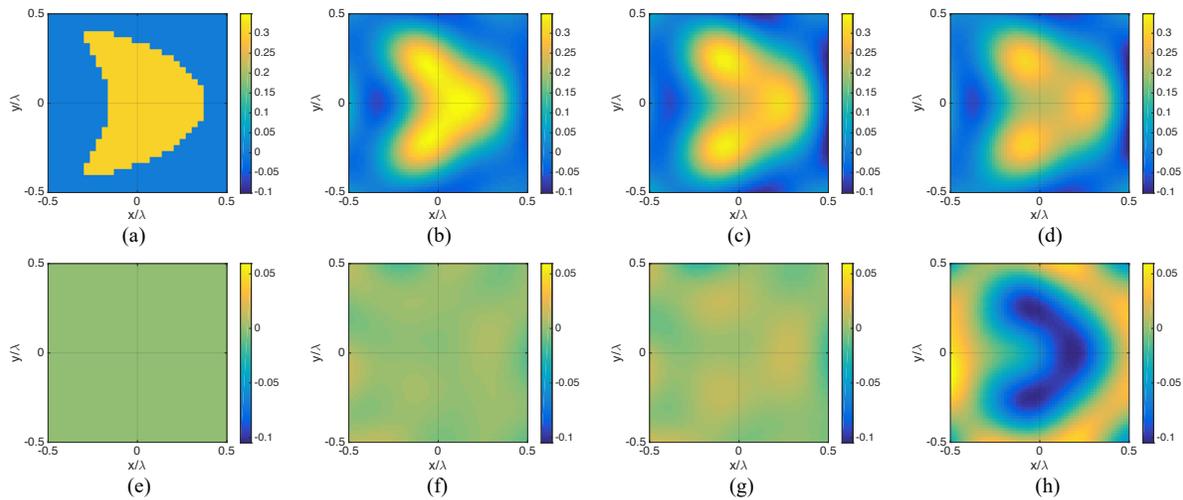

Figure 2. Assessment against numerical data: lossless homogeneous kite target χ=0.3. Real (a) and imaginary (e) parts of the reference profile. Real and imaginary parts of the contrast function retrieved by assuming: known the exact total field (b),(f); by adopting Y0-BA (c),(g) and BA (d),(h).

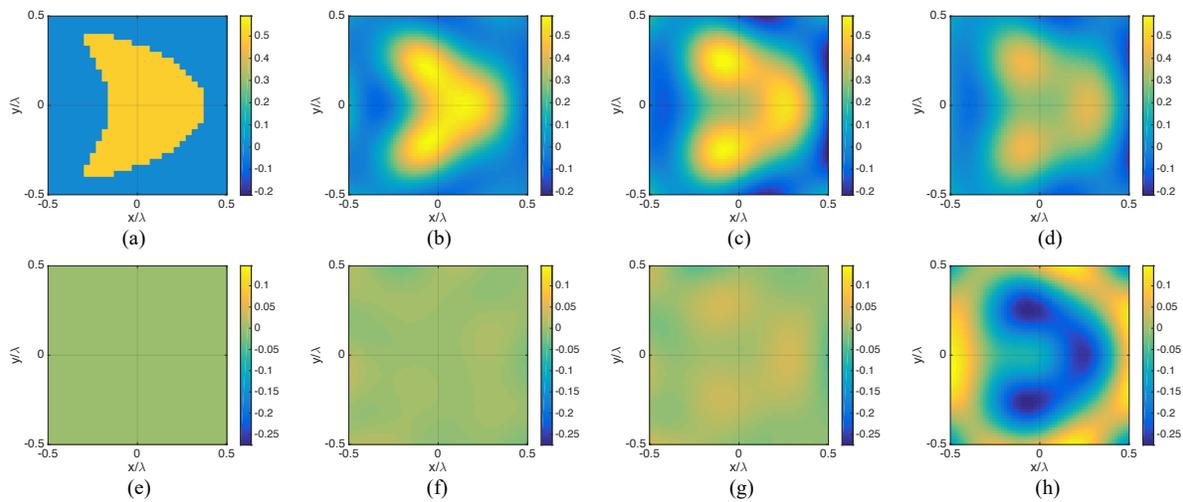

Figure 3. Assessment against numerical data: lossless homogeneous kite target χ=0.5. Real (a) and imaginary (e) parts of the reference profile. Real and imaginary parts of the contrast function retrieved by assuming: known the exact total field (b),(f); by adopting Y0-BA (c),(g) and BA (d),(h).

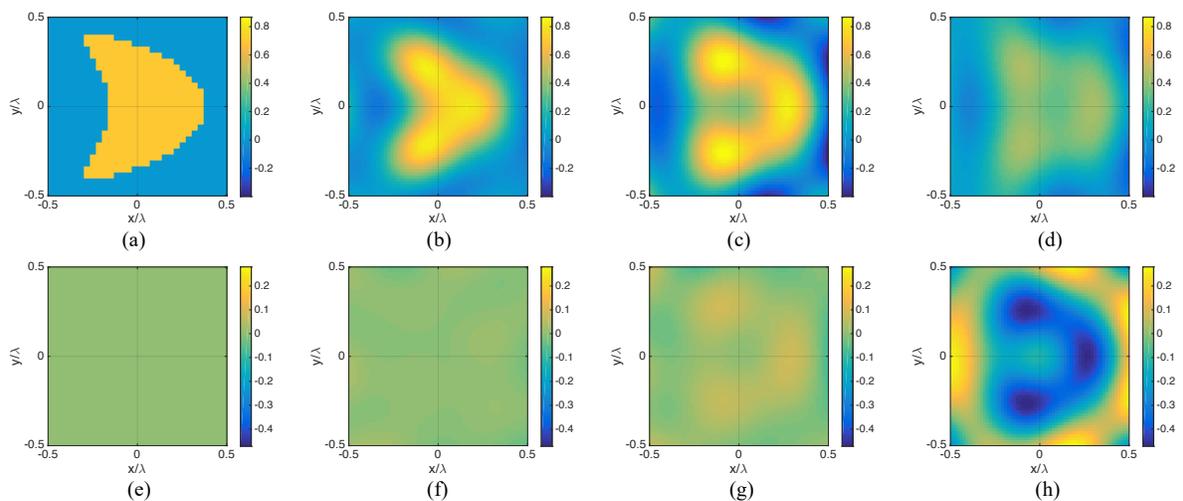

Figure 4. Assessment against numerical data: lossless homogeneous kite target χ=0.7. Real (a) and imaginary (e) parts of the reference profile. Real and imaginary parts of the contrast function retrieved by assuming: known the exact total field (b),(f); by adopting Y0-BA (c),(g) and BA (d),(h).

The results are reported in Figure 6. As can be seen, BA is not able to quantitatively retrieves the cylinders. As a matter of fact, it can only detect them and retrieve their support. On the other hand, Y0-BA can accurately retrieve both the geometrical and the electromagnetic properties of the considered targets.

| | *Ideal case* | **Y0-BA** | **BA** |
|---|---|---|---|
| $\chi$=0.3 | 0.13 | 0.17 | 0.26 |
| $\chi$=0.5 | 0.13 | 0.22 | 0.43 |
| $\chi$=0.7 | 0.13 | 0.3 | 0.64 |
| $\chi$=1 | 0.13 | 0.63 | 0.96 |

Table I. Kite target: normalized mean square errors.

## V. NUMERICAL ASSESSMENT: NON-LINEAR INVERSION

Encouraged by the interesting results achieved by means of Y0-BA, we have tested the Y0 model within a non-linear regime. In particular, the CSI method [9],[20] has been adopted to solve the relevant inverse scattering problem.

The CSI method tackles the inverse scattering problem in its full non-linearity, by contemporarily looking for both the contrast $\chi$ and the auxiliary unknown $W$. In particular, the problem's solution is iteratively built by minimizing a cost functional, which takes into account both the misfit in the data and state equations [9].

In the proposed Y0 model, the standard state equation has been substituted by equation (10). Accordingly, from a mathematical point of view, the new CSI method, that in the rest of the paper is referred as Y0-CSI, amounts at retrieving the unknowns of the problem by minimizing the following cost functional:

$$\Phi(W,\chi) = \sum_v \frac{\left\| \chi(\underline{r})\hat{E}_i^v(\underline{r}) + \chi(\underline{r})A_i^{Y_0}(W^v) - W^v(\underline{r}) \right\|^2}{\left\| \hat{E}_i^v(\underline{r}) \right\|^2}$$
$$+ \sum_v \frac{\left\| E_s^v(\underline{r}_m) - A_e(W^v) \right\|^2}{\left\| E_s^v(\underline{r}_m) \right\|^2}$$
(13)

Due to the lower DNL of the model, a faster convergence and/or a more reliability of the solutions are expected. In summary, the proposed Y0-CSI involves the following three steps:

- evaluation of $E_{J_0}^{\ v}$ from the data $E_s^v(\underline{r}_m)$ via identity (7);
- redefinition of the internal radiation operator and the incident field;
- minimization of the cost functional in (13) according to some optimization procedure.

In particular, in this paper, the optimization of the cost functional in (13) is pursued within a conjugate gradient scheme. Moreover, the initial guess is given by the back-propagation solution. More details can be found in [9].

In order to test and assess Y0-CSI, a comparison has been performed with the results obtained by means of standard CSI method, in the following referred as H02-CSI. In case of simulated data, no additional regularization technique has been adopted. On the other hand, in case of experimental data, the cost function has been equipped with an additive regularization term. In particular, a total variation regularization has been considered by adding to the functional (13) a penalty term, which aims at enforcing a piecewise constant target, i.e:

$$\Phi_P(\chi) = \frac{k^2}{2}\left\| \eta(\underline{r})\mathcal{D}_h[\chi(\underline{r})] \right\|^2 + \frac{k^2}{2}\left\| \eta(\underline{r})\mathcal{D}_v[\chi(\underline{r})] \right\|^2$$
(14)

wherein $k = N_c^{-1}$, $\mathcal{D}_h$ and $\mathcal{D}_v$ represent the partial derivatives with respect to the horizontal and vertical coordinates of the reference system, respectively. $\eta(\underline{r})$ is a weight function which normalizes $\mathcal{D}_h[\chi]$ at a given iteration with respect to the one evaluated at the previous iteration [33].

### A. Simulated data

In the following numerical tests, the same targets as the ones in the Section IV have been considered.

For the kite target, the following parameters have been considered: a contrast value equal to 1-0.6j, L = $2\lambda_b$, 20 receivers and transmitters, R = 2 $\lambda_b$, a working frequency of 300 MHz, SNR = 20dB and $N_c = 60 \times 60$. As far as the Austria profile, a contrast of 1 has been considered with L = $2\lambda_b$, $N_c = 64 \times 64$, 18 receivers and transmitter, R = 4 $\lambda_b$, SNR=20 and a working frequency of 400 MHz.

The results are reported in Figure 7. As far as the kite target, the means square errors are 0.25 and 0.55 when Y0-CSI and H02-CSI has been adopted, respectively. As can be seen, the H02-CSI completely fails in retrieving the permittivity of the target, and overestimates the imaginary part. On the other hand, in case of Austria target, the new CSI method allows one to retrieve the properties of the target with a $NMSE = 0.17$. The results obtained by means of H02-CSI are not shown as it completely fails in retrieving the target ($NMSE = 1$).

### B. Experimental data

In this subsection, the *TwinDielTM* target at a working frequency of 6 GHz has been considered. The region of interest of 0.15 × 0.15 $m^2$ have been discretized in 64 × 64 cells, and a 18 × 18 multiview-multistatic data matrix has been processed.

The results, when the Y0-CSI has been adopted, are shown in Figure 7(g)-(h). As can be seen, both the shape and electromagnetic properties of the targets are accurately retrieved. On the other hand, the H02-CSI method is not able to retrieve the target. As a consequence, the retrieved profile is not shown.

## VI. CONCLUSION

The mathematical model adopted for the solution of inverse scattering problems can significantly condition the reliability and achievable performance of a given inversion technique. In this paper, with reference to 2D scalar problem, a new model for the solution of inverse scattering problem is introduced and tested against both numerical and experimental single frequency data. In particular, the new model, referred as Y0 model, is derived from a convenient decomposition of the Green's function and of the internal radiation operator. Such decomposition has allowed to rewrite the state equation by extracting a term of the radiation operator which is related to the radiating currents and can be easily computed from the data. The above decomposition has led to a redefinition of the relevant incident field and integral operator.

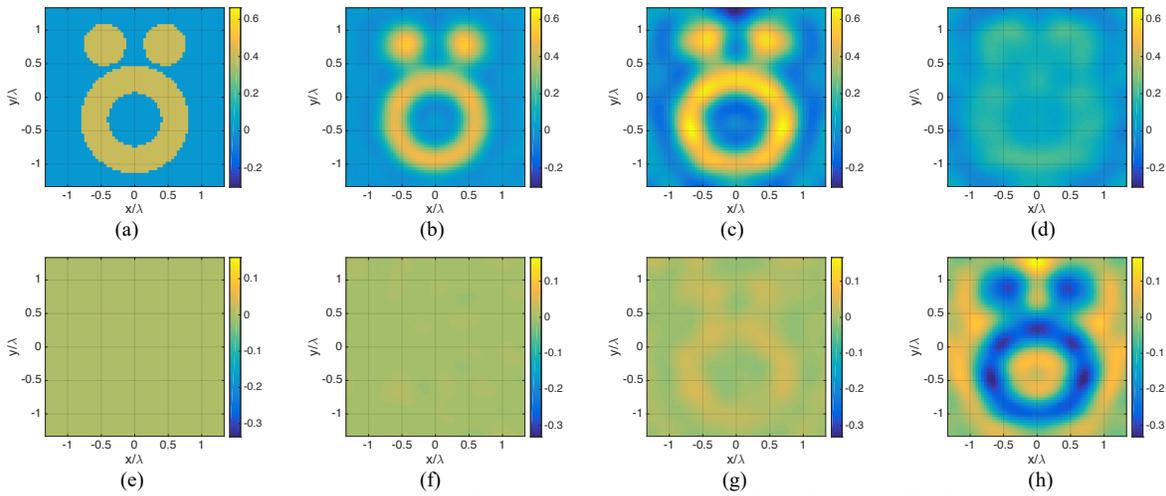

Figure 5 Assessment against numerical data: Austria target $\chi$=0.4. Real (a) and imaginary (e) parts of the reference profile. Real and imaginary parts of the contrast function retrieved by assuming: known the exact total field (b),(f); by adopting Y0-BA (c),(g) and BA (d),(h).

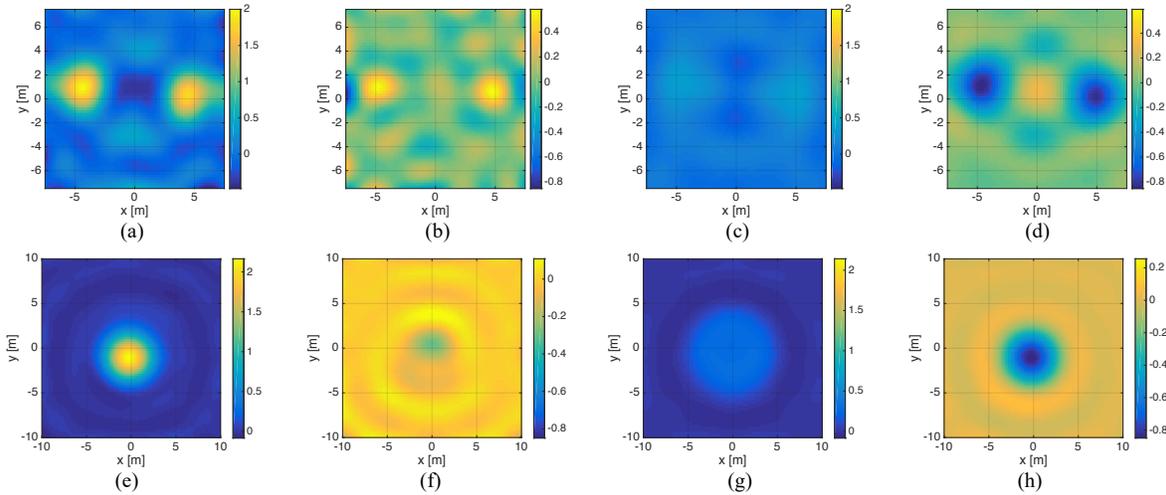

Figure 6. Assessment against experimental data. FoamDielIntTM Fresnel target at 3GHz (a)-(d) and TwinDielTM Fresnel target at 4GHz (e)-(h). Real and imaginary parts of contrast functions retrieved by adopting Y0-BA (a),(b) and (e),(f), and BA (c),(d) and (g),(h).

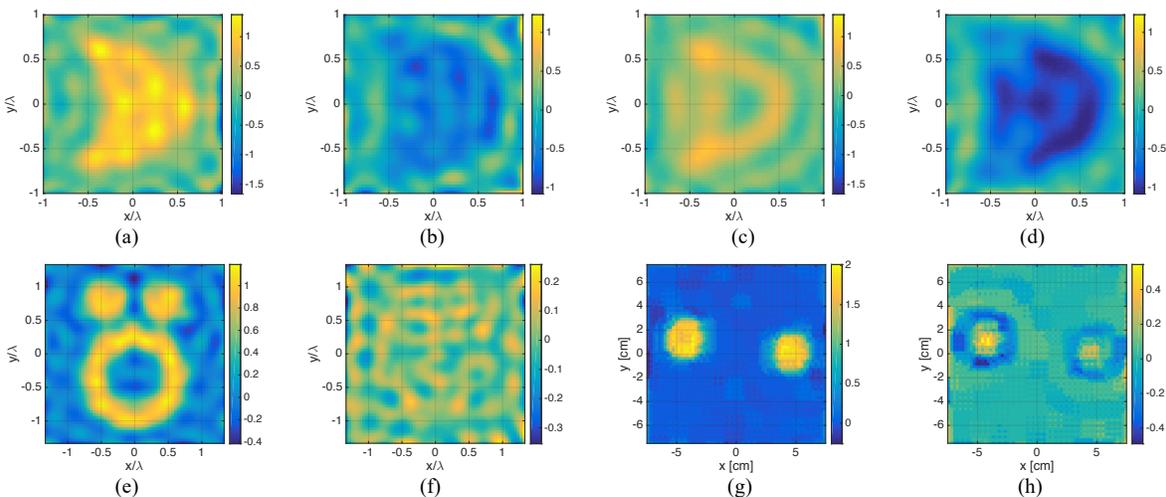

Figure 7. Non-linear framework. Assessment against numerical (a)-(f) and experimental (g)-(h) data. Real (a),(e),(g) and imaginary (b),(f),(h) parts of contrast functions retrieved by adopting Y0-CSI. Real (c) and imaginary (d) parts of the contrast function retrieved by adopting H02-CSI.

The thus obtained model has been proved to have a lower degree of non-linearity as compared to the classical scattering model by means of a comparison of the norms of the two relevant integral operators. Moreover, a new linear approximation of the total field inside the investigation domain has been derived from the new model which definitely outperforms the corresponding linear approximation (i.e., the Born approximation) of the H02 model and exhibit a wider range of applicability. Finally, a modified version of CSI method has been proposed starting from the Y0 model. Due to the lower DNL of the model, faster convergence and more reliable solutions can be obtained, as witnessed by both numerical and experimental data inversions.

Future work will be focused at analyzing actual perspectives of the new model to the more cumbersome case of lossy and/or partially known scenario, as well as to the case of three-dimensional geometries.

## REFERENCES

[1] D. Colton and R. Kress, "Inverse Acoustic and Electromagnetic Scattering Theory", *Springer-Verlag,* Berlin, Germany, 1998.

[2] M. Bertero and P. Boccacci, "Introduction to Inverse Problems in Imaging", *Institute of Physics,* Bristol, UK, 1998

[3] M. Pastorino, " Microwave Imaging", *John Wiley*, New York, May 2010.

[4] R. Persico, Introduction to ground penetrating radar: inverse scattering and data processing. John Wiley & Sons, 2014.

[5] M. Ambrosanio, M. T. Bevacqua, T. Isernia and V. Pascazio, "The Tomographic Approach to Ground-Penetrating Radar for Underground Exploration and Monitoring: A More User-Friendly and Unconventional Method for Subsurface Investigation," in *IEEE Signal Processing Magazine,* vol. 36, no. 4, pp. 62-73, July 2019.

[6] M. Bevacqua, G. Bellizzi, L. Crocco, T. Isernia, "A Method for Quantitative Imaging of Electrical Properties of Human Tissues from Only Amplitude Electromagnetic Data", *Inverse Problem,* vol. 25, n. 2, 2019.

[7] O. M. Bucci, N. Cardace, L. Crocco, and T. Isernia, "Degree of nonlinearity and a new solution procedure in scalar two-dimensional inverse scattering problems," J. Opt. Soc. Am. A, vol. 18, pp. 1832-1843, 2001.

[8] T. Isernia, L. Crocco and M. D'Urso, "New tools and series for forward and inverse scattering problems in lossy media," *IEEE Geoscience and Remote Sensing Letters,* vol. 1, no. 4, pp. 327-331, Oct. 2004.

[9] M. D'Urso, T. Isernia and A. F. Morabito, "On the Solution of 2-D Inverse Scattering Problems via Source-Type Integral Equations," in *IEEE Transactions on Geoscience and Remote Sensing,* vol. 48, no. 3, pp. 1186-1198, March 2010.

[10] T. Isernia, V. Pascazio, and R. Pierri. "On the local minima in a tomographic imaging technique", *IEEE Trans. Geosci. Remote Sens.,* 39:1596-1607, 2001.

[11] Y. Zhong, M. Lambert, D. Lesselier and X. Chen, "A New Integral Equation Method to Solve Highly Nonlinear Inverse Scattering Problems," in *IEEE Transactions on Antennas and Propagation,* vol. 64, no. 5, pp. 1788-1799, May 2016.

[12] M. T. Bevacqua and T. Isernia, "A convenient rewriting to the 2D inverse scattering problem based on the reduced scattered field," 2019 IEEE International Symposium on Antennas and Propagation and USNC-URSI Radio Science Meeting, Atlanta, GA, USA, pp. 1013-1014, 2019.

[13] M. T. Bevacqua, T. Isernia, R. Palmeri, M. N. Akinci and L. Crocco, "Physical Insight Unveils New Imaging Capabilities of Orthogonality Sampling Method," in *IEEE Transactions on Antennas and Propagation.*

[14] A. J. Devaney, and E. Wolf, "Radiating and Nonradiating Classical Current Distributions and the Fields They Generate", *Phys. Rev. D,* vol. 8, n.4, 1973.

[15] E. A. Marengo and R. W. Ziolkowski, "Nonradiating and minimum energy sources and their fields: generalized source inversion theory and applications," in *IEEE Transactions on Antennas and Propagation,* vol. 48, no. 10, pp. 1553-1562, Oct. 2000.

[16] X. Chen, "Subspace-Based Optimization Method for Solving Inverse-Scattering Problems," in *IEEE Transactions on Geoscience and Remote Sensing,* vol. 48, no. 1, pp. 42-49, Jan. 2010.

[17] T. Yin, Z. Wei and X. Chen, "Non-iterative Methods Based on Singular Value Decomposition for Inverse Scattering Problems," in *IEEE Transactions on Antennas and Propagation.*

[18] E. A. Marengo, E. S. Galagarza, and R. Solimene, "Data-driven linearizing approach in inverse scattering," J. Opt. Soc. Am. A, vol. 34, pp. 1561-1576, 2017.

[19] A. J. Devaney, "Mathematical Foundations of Imaging, Tomography and Wavefield Inversion", *Cambridge UK: Cambridge University Press,* 2012.

[20] Van den Berg P. M. and Kleinman R. E., "A contrast source inversion method," Inv. Prob., vol. 13, pp. 1607–1620, 1997.

[21] A. Brancaccio, V. Pascazio, and R. Pierri, A quadratic model for inverse profiling: the one-dimensional case, in *Journal of Electromagnetic Waves and Applications,* 9:5-6, 673-696, 1995.

[22] W. C. Chew, "Waves and fields in inhomogeneous media". *IEEE press,* 1995.

[23] C. J. Sheppard, S. S. Kou, and J. Lin, The Green-function transform and wave propagation, *Frontiers in Physics,* 2(67), 2014

[24] M. Abramowitz and I. A. Stegun, Handbook of Mathematical Functions with Formulas, Graphs, and Mathematical Tables. NewYork, NY, USA: Dover, 1964

[25] M. T. Bevacqua and T. Isernia, "On Spectral Content of Radiating Components of Electromagnetic Sources," *2019 International Conference on Electromagnetics in Advanced Applications (ICEAA),* Granada, Spain, 2019, pp. 0917-0919.

[26] E. A. Marengo and R. W. Ziolkowski, "Non-radiating and minimum energy sources and their fields: generalized source inversion theory and applications," *in IEEE Transactions on Antennas and Propagation,* vol. 48, no. 10, pp. 1553-1562, 2000.

[27] M. N. Akıncı, "Improving Near-Field Orthogonality Sampling Method for Qualitative Microwave Imaging," in *IEEE Transactions on Antennas and Propagation,* vol. 66, no. 10, pp. 5475-5484, Oct. 2018.

[28] L. Crocco, L. Di Donato, I. Catapano and T. Isernia, "An Improved Simple Method for Imaging the Shape of Complex Targets," in *IEEE Transactions on Antennas and Propagation,* vol. 61, no. 2, pp. 843-851, Feb. 2013.

[29] O. M. Bucci and T. Isernia, "Electromagnetic inverse scattering: Retrievable nformation and measurement strategies", *Radio Sci.,* 32:2123–2138, 1997.

[30] J. Richmond, "Scattering by a dielectric cylinder of arbitrary cross section shape", *IEEE Trans. Antennas Propag.,* 13(3):334-341, 1965.

[31] K. Belkebir and M. Saillard, "Special section: Testing inversion algorithms against experimental data", *Inverse Probl.,* 17:1565–2028, 2001.

[32] K. Belkebir and M. Saillard, "Special section: Testing inversion algorithms against real data: Inhomogeneous targets", *Inverse Probl.,* 21, 2005.

[33] P. M. van den Berg, A. Abubakar, J. T. and Fokkema, Multiplicative regularization for contrast profile inversion, *Radio Sci.,* 38, 8022, doi:10.1029/2001RS002555, 2, 2003.